\documentclass[12pt,a4paper,notitlepage]{article}
\usepackage[dvips]{graphicx}
\usepackage{verbatim}
\usepackage{amsfonts,amssymb,textcomp}

\usepackage{epsfig}
\usepackage{amsmath}
\usepackage{hyperref}

\begin{document}

\thispagestyle{empty}

\title{Technique of production of argon-37\\
        at proton cyclotron\\
and detector for measurements}
\author{A.~D.~Romazanov}
\date{\it Institute of nuclear physics\\ \it Academy of sciences of Uzbekistan\\
\it Toshkent, Ulugbek village, Uzbekistan}

\maketitle

\begin{abstract}
The technology of production of the isotope Ar-37 at proton cyclotron is developed. It is based on irradiation of the Cl-37 target with the protons of energy of a few a MeV. The example of production of tiny amount of Ar-37 is described and discussed. The detectors to measure the intensity of the sample is discussed.
\end{abstract}

\section{Introduction}

The $^{37}$Ar isotope is very important in different fields of physics. It decays by 100\% electron capture (EC) 3/2$^+\rightarrow$3/2$^+$ transition to the ground state of the stable nuclide $^{37}$Cl. The  decay scheme is complete as there no excited levels of $^{37}$Cl below the EC decay energy Q$^+$=813.87~keV \cite{endt:1998}. Due to the monoenergetic neutrino lines (811 and 813~keV) and the absence of the nuclear $\gamma$-radiation it is suitable to test neutrino detectors. In 1988 Haxton proposed to employ the $^{37}$Ar source to calibrate radiochemical detectors of solar neutrinos \cite{haxton:1988}, especially gallium based one (\cite{jna:2009}, \cite{jna:20063}). Approximately 1 MCi source was produced \cite{barsanov:2007} according to that proposal and the SAGE detector was successfully calibrated \cite{jna:20062}. Another perspective possible usage of the isotope is a calibration of low energy electron detectors by means of the low energy Auger electron and X-ray emission. It can be used in the experiments aiming to search for a possible sterile neutrino admixture in the $\beta$-spectra of different isotopes. In particular, the experiments with tritium \cite{jna:2014} are possible employers of the $^{37}$Ar source.

\section{Technology of production by neutron irradiation}

The 1~MCi $^{37}$Ar used for the gallium detector calibration neutrino source was manufactured by irradiating a piece of pressed calcium oxide in the fast breeder reactor BN-600 in Russia \cite{barsanov:2007}. The technology was based on the $^{40}$Ca(n,$\alpha$)$^{37}$Ar reaction. The fast neutron flux was measured by an organic scintillator \cite{fisher:2011}. A special facility was build \cite{barsanov:2006} in order to extract the gaseous $^{37}$Ar sample from the CaO target. An obvious disadvantage of the method is that after neutron irradiation there is a lot of $^{39}$Ar - in \cite{barsanov:2007} they report the contamination to be of 0.34\% of the gas volume fraction. 

\section{Technology of production by proton irradiation}

The reaction $^{37}$Cl(p,n)$^{37}$Ar looks to be able to produce much more pure samples of $^{37}$Ar. One of the first usage of the method was published in \cite{colomer:1973}. We have developed a technique of preparation of the target based on a KCl film. The target is a Nb foil $20{\times}10{\times}1$~mm with a spot of thin KCl film deposited at the foil in vacuum. The spot is roughly circular with the area of 1.5~cm$^2$. The thikness of the film is 100~$\mu$m, the mass of the KCl is about 30~mg. The target was firstly irradiated at the Moscow State University proton cyclotron of the Institute of Nuclear Physics; the energy of protons was $E_p{=}7$~MeV and the current was $I_p{=}2$~$\mu$A.

The calculation of an expected activity of $^{37}$Ar was done without taking into account of ionization energy loss. The maximal expected value of the intensity may be estimated as 
\\
\begin{displaymath}
N=N_p \sigma n = \frac{I_p {\Delta} t}{e} \sigma \rho d \frac{N_A}{A}
\end{displaymath}
\\
Here $N$ is the number of $^{37}$Ar atoms produced in the reaction; $e$ is the electron charge, 1.6${\cdot}10^{19}$~Ql; $N_p$ is the number of protons bombarded the target during irradiation for the time period ${\Delta} t$ and the current $I_p$; $\sigma$ is the cross-section of the reaction; $\rho$, $d$ is the density and thikness of the target; $A$ is the atomic number of the target; $N_A$ is the Avogadro constant, 6.02${\cdot}10^{23}$~$\frac{1}{\textrm{mol}}$.

An isotope $^{37}\mathrm{Ar}$ is produced in the reaction $^{37}$Cl(p,n)$^{37}$Ar that has a cross-section of ${\approx}0.5$~barn at the proton energy $E_p{=}7$~MeV. The target KCl has the molar mass of 74.5~$\frac{\textrm{g}}{\textrm{mol}}$ and the density of 1.98~$\frac{\textrm{g}}{\textrm{cm}^3}$. Taking into account the 24.2\% abundance of $^{37}$Cl in the natural target we may expect for 2~hours irradiation $ N=3.5{\cdot}10^{11} \textrm{\ \ atoms of \ \ } ^{37}\textrm{Ar}$.  Because half-life of $^{37}$Ar is 35~days it corresponds to $5.7{\cdot}10^4$~Bq of the intensity immediately after end of bombardment. The real sample was obtained after 2-hours irradiation and filling a proportional counter at special system \cite{jna:2011}. The proportional counter was made according to the technology  described in \cite{yants:1994}. The result of the intensity measurement was about 5.3$\pm$0.6~Bq that is in good agreement with the calculation.
\\

\section{Conclusion}

The technology of production of pure $^{37}$Ar based on irradiation the KCl target with protons with an energy of a few MeV is developed and tested for small amount of source. The proportional counter is most suitable detector to measure a tiny amount of $^{37}$Ar.

\bibliography{romz}
\bibliographystyle{h-physrev}
\bibliographystyle{hplain}

\end{document}